# Development of Railway Silk Road as a Platform for Promoting Georgia's Economic Growth

Davit Gondauri[1] & Manana Moistsrapishvili[2]

[1] Business and Technology University, Tbilisi, Georgia

[2] Georgian Technical University, Tbilisi, Georgia

Correspondence: Davit Gondauri, Business and Technology University, I. Chavchavadze Avenue N82, Tbilisi 0162, Georgia. E-mail: dgondauri@gmail.com

**Abstract**

The given paper emphasizes on the importance of the Railway Silk Road for promoting the Georgia's economic growth and development. The article notes that the process of economic integration of the region allows the increase of cargo turnover volume in Central Asia and the Caucasus countries, thus increasing the transport of goods through Georgia, - contributing to the sustainability of Georgia's macroeconomic and economic growth. The aim of financially economical models, we have identified the causal links between the sensitivity of railway cargo and the economic growth of the country. The main task of the research was to use the Railway EVA and the Georgian economy, create a cargo sensitivity relationship between CAGR models. The paper presents the analyses of the scientific research problems regarding the Railway freight transportation studies. Calculations are provided regarding the share of the Railway System in the country's GDP in period of 2006-2017 year and the average annual geometric (CAGR) growth pf cargo volumes in 16 years' cycle allowing the „Georgian Railway"JSC to create the additional value in the country's overall GDP. The research indicate that the additional value to GDP is the direct and indirect form of development and growth of various sectors of the economy of Georgia, as some part of the cargo shipped in the railway remains in Georgia and is used in the process of production, which in itself adds value added to the economic growth of the country. Also, the use of this model by scientific research in foreign research centers provides better opportunities for additional growth of their economies.

## 1. Introduction

*1.1 Problem Definition*

The factors such as the: region's economic development, utilization of fossil resources in Central Asia, the economic process of sustainable growth will unavoidably increase the turnout of the cargo transportation in the Middle East and the East Caucasus countries. Important that the Silk Road, a trans-Eurasian network of trade routes connecting East and Southeast Asia to Central Asia, India, Southwest Asia, the Mediterranean, and northern Europe, which flourished from roughly 100 BCE to around 1450, has enjoyed two modern eras of intense academic study (Andrea, 2014, p. 105). Subsequently, the increased volume of freight transported via Georgian corridor will positively support the country's Macroeconomic stability and benefit the Economic growth. The current economic and logistical situation gives the multiple options and alternative ways of transportation rather than the "Georgian Railway" JSC, therefore, to gain the advantage country and the company has to offer to the customers a low-cost railroad transportation, with an increased speed of transportation, improved reliability and more simplified transport service.

## 2. Review of the Literature

Railway System is an important part of Georgia's overall economy. Despite its smaller share in the country's GDP, "Georgian Railway" JSC role in the socio-economic development of the country is far more pronounced.

Table 1. The railway systems Economic share in Country's overall GDP (Note 1)

| Year | Railway Transport | GDP market prices | % in GDP |
|---|---|---|---|
| 2006 | 65.6 | 13,790 | 0.476% |
| 2007 | 57.5 | 16,994 | 0.338% |
| 2008 | 54.0 | 19,075 | 0.283% |
| 2009 | 49.0 | 17,986 | 0.272% |
| 2010 | 62.9 | 20,743 | 0.303% |
| 2011 | 71.1 | 24,344 | 0.292% |
| 2012 | 70.2 | 26,167 | 0.268% |
| 2013 | 70.1 | 26,847 | 0.261% |
| 2014 | 71.0 | 29,150 | 0.243% |
| 2015 | 76.2 | 31,756 | 0.240% |
| 2016 | 59.0 | 34,028 | 0.173% |
| 2017 | 55.0 | 37,847 | 0.145% |

From the examination of Table 1 given above it becomes clear that, in the specified period 2006-2017 the volume of the share in County's GDP created by the railway sector is characterized by decreasing trend, which in its turn, is caused by the correlation of external and internal shocks of the economy.

Amendments made in the political life of Georgia raise significant perspectives for further development of foreign investments. In this regard, one of the most prominent fields in Georgia should be the direction of transport. Its impact on the growth of the economy is very broad and diverse. This is expressed both directly and indirectly. Direct impact implies increase of state and budget revenues, employment, attraction of investments. Indirect impact is expressed in the development of transport in the stimulation of other sectors of economy. Significant improvements have been made in today's transport, especially for railway transport. Recent investments in foreign investments have been attracted in the field of transport (including railways), thus rehabilitation of railway infrastructure and rolling stock. This will in turn ensure safe and quick movement of cargo in the Georgian Rail Corridor:

• 2015 - 608.1 million USD (36.5% of the total investment);

• 2016 - 576,04 million USD (36.8% of the total investment);

• 2017 - 472.6 million US dollars (24.9% in total investment);

• 2018 - 209.9 million US dollars (17% in total investment). (Note 2)

For example, one of the research aimed to study the impact of 'One Belt, One Road' and its effects on GRP growth in Chinese provinces affected by the initiative. The calculated forecasts in this study indicate on decreased GRP growth in coming years. In order to prove that 'One Belt, One Road' and its infrastructure projects will have an impact on GRP, regressions analyses, including several variables that have an effect on GRP (Ylander 2017). Other research show the Chinese New Silk Road (NSR) initiative seems to be the greatest logistics endeavor of our times. It will boost infrastructure investments and create many new relations. The primary aim of the study is to determine the factors influencing the development of the rail part of the NSR. The methodology of the paper includes a desk research about the NSR potential and identification of STEEPVL analysis factors shaping the NSR. The factors are grouped into seven dimensions: social, technological, economic, ecological, and political, and related to values and legal aspects. They constitute factors that can either enhance, accelerate or hamper the success of the NSR. The outcomes of the study comprise the introduction to the complete STEEPVL analysis (Nazarko & Kuźmicz, 2017, 497–503).

How will railroad freight be affected if the annual U.S. GDP growth breaks the three per cent mark? Petr Ledvina, Economist at the Railroad Tie Association (RTA) writes that an increase in U.S. railroad freight traffic is likely should the country's annual GDP growth break the three per cent mark (Note 3).

Railway transportation is playing an extremely important role in promoting the development of regional economy. The research aims at exploring the inner link between regional railway freight volume and regional GDP with the statistical method by exploring the relationship between changes of railway freight volume and regional GDP. Using Cluster Analysis, this paper studies the commonness of the development of regional GDP and railway traffic volume. At last, a proposal is offered for the balanced development of railway and regional economic (Zhenji Zhang, Zuojun Max Shen, Juliang Zhang, Runtong Zhang 2014).

The rail industry in Great Britain and its supply chain employ 212,000 people, generating £9.3bn of gross value added (GVA) a year. The rail sector returns £3.9bn in tax to the Exchequer a year, offsetting nearly all of the £4bn that it receives in government funding. The sector contributes considerable enduring benefits to the productive potential of the economy, including alleviating congestion in the road network and facilitating the development of clusters of economic activity. Oxera has valued these benefits at up to £10.2bn a year. In addition to the economic impacts, the rail sector delivers significant environmental and social benefits. The sector reduces $CO_2$ emissions by up to 7.4m tonnes, valued at £430m annually. It also lowers the numbers of people killed or seriously injured on the transport network in Great Britain (relative to what would be expected to happen if those journeys were made by car or lorry) by up to 950 a year, valued at £330m annually. The sector has been highly resilient in the recent recession, performing better than the rail sectors in other large European countries (despite the UK experiencing a more significant reduction in national gross domestic product, GDP), and better than in previous recessions. (Note 4)

Tashguzar–Boysun–Kumkurgon railway line in Uzbekistan encouraged an increase of around 2% in the regional gross domestic product growth rate in affected regions in the frame of connectivity effects. This seems to have been driven by increases in industry value added and services value added of approximately 5% and 7%, respectively. Positive and significant changes in the industrial output of the directly affected and neighboring regions mostly took place during the design and construction period in anticipation of the railway connection. The impact on agricultural output has been moderate in comparison to the abovementioned sectors, constituting around 1%, which is consistent with previous literature on the differential impact of public capital. Our results and the framework provided might help regulatory bodies to conduct comprehensive estimations of the impact of infrastructure and develop the formulation of both promotional and compensatory measures related to or induced by the effects of infrastructure provision (Yoshino, Naoyuki; Abidhadjaev, Umid 2015).

Model is used to analyze the logistic connectivity of China's 31 provinces by focusing on 11 variables, including some new factors (Density of road network, Density of railway network, Number of Internet Users) not used in previous studies, over the 13-year period from 2002 to 2014. Using panel data regression analysis, the empirical results show a statistically significant and positive impact of transport connectivity (factors like Density of road network, Density of railway network and Number of Internet Users) on economic development in China. In particular, the Number of internet users is a key factor reflecting information connectivity in all the variables. Comparative analysis regarding economic development is conducted to benchmark between coastal provinces and interior provinces. Like most previous research, this study yields the same finding of higher impact of transport connectivity on economic development in eastern provinces than in western provinces. This study suggests that decentralized decision-making will be significantly more efficient for analyzing regional infrastructure development. It also shows that the influence of transport connectivity on economic development is dependent on a certain developmental stage. This suggests that an economic region should adopt different development strategies for transport connectivity during different stages of development (Kevin X. Li, Guanqiu Qi, 2016).

In this regard, several scientists have presented the research papers based on the problems of the railway transportation. Must be noted that lack of literature was a weakness of research. Also, the literature presented on this topic is non-systematic. In this article we will touch the works of Georgian researchers:

One of the Doctoral Studies about Modern Conditions of Transport Management Processes and its Development Perspectives, presents the technical-economic results of the railway industry in the past years The total length of railway track network is 1575 km, 67.4% of which is single track line and mostly electrified. On the different sections of the railway network the Various speed limitations are introduced, approximately on 11% of the total railway network. It should be noted that the rolling stock of the "Georgian Railway" JSC was much more amortized and consequently this factor resulted in the increase of costs for its exploitation. Nearly 70% of the current cargo shipments consisted of liquid cargo, and only a third of the wagon depot park consisted of cisterns. Therefore, there is a significant imbalance between those two categories. Lately, the share of the transit freight transportation is decreases in the overall volume of cargo transportation.

Unfortunately, this fact cannot be considered as a positive event, because as it has been noted there is a high correlation of external and internal shocks of the economy in relations with the transit cargo transportation (Note 5). The Georgian Railway is an infrastructural company designed to transport cargo from Central Asia to Europe and vice versa. When these regions are observed to reduce economic activity, this will be reflected in the shipment of cargo from one point to another and vice versa, which in itself reflects the sensitivity of the railway turnover. However, in the nearest future, the consumption of the fossil wealth in Central Asia and the sustained growth of the economy will increase the volume of the freight turnover in Central Asia and the Caucasus countries. Consequently, the volume of the Cargo transported through Georgia will increase accordingly (A. Kurtanidze 2012, p. 39).

The formation of Georgia as an independent sovereign state and the transition to the market economy automatically assigned „Georgian Railway" JSC the status of the transit road, and now it has become the sole transport link between Asia and Europe. Although the general direction of the railway operation system remains unchanged in the market economy (The Utilization of maximum transportation capacity with the minimum operating costs), the main tasks of railway transport have been drastically altered.

As a result of the European Union's active efforts in 1993, by the power of Brussels Declaration, Georgia, Azerbaijan, Armenia, Kazakhstan, Uzbekistan, Turkmenistan and Tajikistan have confirmed the relevance of the Europe-Caucasus-Asia transport corridor and the intention to solve the international shipping and trade regulation issues respecting the generally accepted international rules and conventions respectively. This corridor, where rail cargo transportation is carried out from Central Asia to Europe and vice versa, includes the following route - Central Asia region - Caspian Sea, Georgia's railway infrastructure - Black Sea - Europe. Due to this, the transportation of one cargo from one point to the other requires a railway as well as marine (ferry) transportation. Ferry transportation in international traffic ensures more rapid and economical delivery of cargo to its final destination; The timing for the shipping of the freight ships is significantly reduced in ports, thus decreasing the cost of the goods from one type of transport to the other, at the same time increasing the reliability and security of the freight and reducing the number of the damages and losses in reset operations. In addition to all the above mentioned the labor costs for ship preparation decreases together with the expenses on the construction, mechanization and storage of warehouses. Taking into account all the above mentioned the ferry passages of Batumi and Poti justify their designation.

The European Union considers this regional technical assistance project (TRACECA) in relations with its global strategic purpose:

1. By implementing this project EU wants to support the strengthening of the sovereignty of the South Caucasus and Central Asian countries, first of all Georgia, Azerbaijan, Armenia, Kazakhstan, Uzbekistan, Kyrgyzstan, Tajikistan, Turkmenistan, and their political and economic independence in the post-Soviet space. By means of linking the European and world markets through alternative transport-transit communication facilities. Implementation of this idea will positively change the outlook of any Eurasian country, including Georgia.

2. The EU wants to create TRACECA to promote regional collaboration and co-operation in the Southern Caucasus and Central Asia. This is undoubtedly a positive project and Georgia will undoubtedly support its proper implementation.

3. The TRACECA project will increase global investments in South Caucasus and Central Asian countries by using the help from international financial institutions as well as using private investment resources. This process is vital for the development and growth of Georgia and the entire region.

4. One of the main strategic goals of the project is to link the Eurasian Transport Corridor, with the Trans European corridor and therefore with the world`s transport communications.

Taking into account all the above mentioned, it is clear that the Eurasian Transport Corridor is a multifunctional transport communication that will provide the serviced necessary for the export of rich resources from the Caspian Sea region as well as from Russia, Central Asia and Kazakhstan together with the other South Caucasus countries.

Thus, with the aim of financially-economical models, we have identified the causal links between the sensitivity of railway cargo and the economic growth of the country. The main task of the research was to use the Railway EVA and the Georgian economy, create a cargo sensitivity relationship between CAGR model.

Empirical research methodology alone accounted for 71 percent among all four methodologies (empirical, conceptual, descriptive and exploratory cross-sectional). From the analysis of studies, role of accounting adjustments, empirical evidences in developed economies, EVA as a strategy, EVA and discounting techniques like NPV, IRR and managerial performance measurement aspects of EVA (Sharma & Kumar, 2010). Thus EVA determines the criteria of business performance, the effectiveness of its financial structure, as well as a single reference rate for the various activities of the company - financial activities, investment activities etc. The progressing globalization and the development of international economic relations are gradually enforcing not only harmonized accounting but also unified theoretical and methodological basis for the assessment of individual aspects of the company and its operations (Salaga et al., 2015).

## 3. Research Methods

In the process of the analyses of the actual materials and the tasks set out in the article uses the empirical data systematization, verbal discussion of their results, conclusions and general regularities of the relationship between the formal - logical description, statistical research methods, classification methods, system approach, forecasting, statistical data monitoring, comparative analysis, generalization and systematization methods.

The research was used by the volume of the freight carried out in the 2003-2017 region of the Georgian Railway. Data was taken from the Georgian Railway Information Technology Agency. We observed the sensitivity of cargo movement in the region. Systematization of data from Georgian Railway Information Technologies Service and OSJD Information Source and Comparative Analysis. From the statistical research methods, we have chosen to analyze the regression model. Classification of cargo shipments in the region according to their categories. In the study with regressive analysis, the company's (Georgian Railway) added economic growth model EVA (Note 6), financial modeling methods NPV (Note 7) and CAPM (Note 8).

The matrix was calculated in the Rail Operation 16-year cycle of cargo volume increase by CAGR, which changes 1-15% to create a regression model. To calculate EVA, we have selected data from the Georgian Railway Financial Statements - the data of the profit-loss and balance sheet. In the regression model for EVA and GDP correlation the GDP is taken from the official statistics of the Georgian statistics service that is subdivided into the web page. In the regression model for EVA and GDP correlation, the discounted rate of cash acceptable cash flow for the future of the railway business is used - the Capital Asset Pricing Model uses the count.

So that the CAPM with more realistic assumptions, it completes its original framework by including (1) risk-taking investors in the investor population, (2) investors who can have heterogeneous expectations or beliefs – an overlooked but required condition for the CAPM to be an internally consistent and meaningful model of competitive financial asset pricing under uncertainty and (3) a positive-sloped short-run supply curve based on a reasonable interpretation of the nature of financial asset trade (Dawson, 2014).

During the implementation of the research we have used the methods of regressive analysis. Together with the mathematical approaches to correlate density, we have formulated the general equation of the railway silk road effectiveness. It is based on factor analysis based on the methods of extrapolation, for which various mathematical and economic methods are used.

CAGR is one of the most accurate ways to calculate the return on an investment that rises and falls in value during the investment period. CAGR allows investors to compare investments with different time horizons. CAGR makes it possible to compare profits from a particular investment with risk-free instruments. It also allows you to assess whether the premium for the risk taken is high enough. The main advantage of using EVA as a metric for performance appraisal is that it takes into consideration all the costs including the cost of equity capital, which is ignored in normal accounting. With this EVA Model, economic profit can be determined.

External and internal factors acting on the results of the research (attracting the cargo):

1. The growing tensions between Russia and the European Union and between Russia and Turkey, is forcing China not to depend on Russia's only railway route. Besides, the political factor there is another problem related to the overcrowding of the railway infrastructure.

2. In the Trans-Asian Railway network countries such as: Kazakhstan, Azerbaijan, Georgia and Turkey are interested in combining Caucasian (trans-Caspian) railway sections. Attraction of China will be possible if we join the "Viking" container trains, and Ukraine - Georgia will replace ferryboats by the container vessels on the Black Sea.

3. After the deteriorating conflict with Russia due to Syria, Turkey is trying to shift the cargo on Baku-Tbilisi-Kars railway and Bosphorus strait. The Baku - Tbilisi - Karsi axis can be used for the transportation of Russian - Turkish cargo shipments with an additional 1.5 million tons of wheat.

4. The Railway corridor "India - Persian Gulf - Iran - Azerbaijan - Georgia - Ukraine - Europe" significantly reduces the time for the cargo shipment. For example, while currently, time necessary for carrying freight from India and Persia to Europe is approximate 30-40 days, the number of days in the new railway corridor will be reduced Down to 15.

5. The "Georgian Railway'' JSC has a minimum tariff when transporting cargo through "South-West" route.

6. In case of complete modernization of the Georgian Railway line sharp rise in the transit transportation is expected. After completing the modernization project the Georgian Railway network will increase its freight transportation capacity and reach 45,055 million tons per year. The increased capacity will be a result of full Modernization of the railway track and supporting infrastructure with the cost of the project approximately 270 million Swiss francs. With the increase of the capacity, the special emphasis should be made on the development of the terminal network for the cargo ports.

7. The construction and development of the Anaklia Deep Sea port will allow Georgian Railway JSC to attract the increased amount of the cargo in the future, the extent of Anaklia`s cargo spreads are supposedly the Transcaucasian countries such as: Kazakhstan, Turkmenistan, Uzbekistan, Tajikistan and Kyrgyzstan. The above mentioned countries are expected to play the role in distribution of the freight coming from China. The Anaklia Deep Sea Port has already concluded the agreements for the cargo of 3 million tons of dry goods, bulk cargo and container flows, and the shipment process will begin as soon as the first phase of the development is completed.

8. The Caspian region is rich with high quality oil fields such as: Buzachi, Kumkol, Alacja, Okarem, Chelekemi and many others. The owners of those high quality fields are not interested in mixing their product with cheaper materials during the realization of the product, even with the Russian Oil so called Ural's Sort. For that purpose, the producers prefer to ship their valued freight via railway instead of the pipeline. This oil has been actively moving through the Georgian Railways network and Terminals, and even today there is still the oil industry's interest in Georgian corridor as an alternative route for the shipment of the cargo. Therefore, in given situation, Georgia would greatly benefit if the subject will be regulated at the level of the state policy in order to use the opportunity and gain the shipments from Turkmen oil producers such as Chelekem oil field (shipped through Iranian Necka Port 60-70 tons per month) via "Georgian Railway" JSC.

9. Turkmenistan Factor when attracting the cargo –Soil, Clay and Aluminum (50% discount on transit transports in Tajikistan via Port of Turkmenistan) for Uzbekistan; Turkmenistan for transiting freight such as oil and gas condensates with 30% discount, 50% discount on sulfur compartment and 30%off from sulfur transit, both products are in high demand in Turkish market. Consequently, the most rational and convenient way of transporting those goods go through Baku-Tbilisi-Kars railway (in total 2-3 mln tons of freight).

Table 2. The volume of cargo transported by the ''Georgian Railway'' in 2003-2017. These data are taken from the internal corporate website of the Georgian Railway Information Technology Agency

| | 2003 | 2004 | 2005 | 2006 | 2007 | 2008 | 2009 | 2010 | 2011 | 2012 | 2013 | 2014 | 2015 | 2016 | 2017 |
|---|---|---|---|---|---|---|---|---|---|---|---|---|---|---|---|
| Volume of transported fright | 16558.6 | 15424.4 | 18986.8 | 22643.3 | 22231.4 | 21181.2 | 17104 | 19930.1 | 20123.3 | 20076 | 18185 | 16673.3 | 14142.7 | 11881.6 | 10672.6 |
| Among them: local | 2410.1 | 2089.1 | 2109 | 2531.2 | 2852.3 | 2406.1 | 1708.2 | 2203.7 | 3009.4 | 2974.4 | 2525.6 | 2572.9 | 2393.1 | 1982.4 | 1929.8 |
| Import | 1057.4 | 1675.5 | 2119.6 | 2663.5 | 3562.2 | 3477 | 2639.4 | 3034.9 | 2932.2 | 3274.7 | 2744.7 | 2966.1 | 2691.2 | 2643.3 | 2663 |
| Export | 786.2 | 919.8 | 1228.8 | 1267.6 | 1404.5 | 1498 | 1380.6 | 1473.4 | 1598 | 1711.5 | 1740.9 | 1625 | 1109.9 | 1073 | 1103.2 |
| Transit | 12305 | 10739.9 | 13528 | 16180.6 | 14412.2 | 13799.4 | 11375.8 | 13218 | 12583.8 | 12115.4 | 11373.8 | 9509.4 | 7948.5 | 6183 | 4976.5 |
| Oil and related products | 10513.8 | 8659.3 | 11412.4 | 13539.5 | 11531.7 | 10062.7 | 9727.5 | 11456.1 | 10459.4 | 9470.7 | 9088 | 7514.4 | 6744.4 | 5494.1 | 4347.6 |
| Among them: Crude oil | 5657.9 | 4010.5 | 6544.1 | 8540.7 | 6523.1 | 4907.6 | 5181.2 | 6327.4 | 5399.6 | 4718 | 3956.9 | 1676.2 | 863.8 | 1808.3 | 401.5 |
| Dry goods | 6044.9 | 6765.1 | 7574.4 | 9103.8 | 10699.7 | 11118.4 | 7376.4 | 8474 | 9664 | 10605.4 | 9097 | 9158.9 | 7398.3 | 6387.5 | 6325 |
| Among them: Aluminum oxide | 253.7 | 303.1 | 372.4 | 426 | 428.5 | 317.5 | 319.7 | 368.2 | 296.7 | 406.8 | 510.7 | 325.5 | 145.7 | 125.7 | 117.4 |
| Boxi | 558.5 | 818.9 | 1067.4 | 1225.1 | 753.5 | 904.6 | 39.3 | 0 | 88.9 | 294.5 | 0 | 0 | 0 | 0 | 0 |
| Black metal | 514 | 575.9 | 377.1 | 440.1 | 602.7 | 646 | 463 | 715 | 862.5 | 905.1 | 883 | 1033.2 | 882.7 | 665.4 | 532.2 |
| Black metal scrap | 915.1 | 465.3 | 296.8 | 407.6 | 420.1 | 362 | 293.7 | 320.1 | 302.2 | 199.6 | 48.2 | 30.4 | 10.3 | 4.1 | 0.9 |
| Industrial raw materials | 148.4 | 539.4 | 845.5 | 525.3 | 913.1 | 1151 | 330.3 | 272.6 | 220.4 | 432.8 | 429.8 | 276.3 | 179.2 | 194.3 | 240.3 |
| Construction materials | 944.3 | 603.6 | 652.8 | 1229.4 | 1783.8 | 1599.8 | 1032.8 | 1163.8 | 1617.5 | 1554.7 | 1376.4 | 1750.3 | 1372.1 | 1019.9 | 1125.6 |
| Wheat and wheat products | 549.4 | 923.7 | 747.6 | 1067.9 | 1110.9 | 794.2 | 1015.7 | 1269.5 | 1153 | 1377.3 | 889.1 | 825.8 | 686.5 | 418.5 | 284.5 |
| Sugar | 499.9 | 506.6 | 617.9 | 522.5 | 683.2 | 563.3 | 513 | 602.8 | 533.9 | 689.3 | 610.1 | 617.5 | 464.2 | 499.5 | 378.5 |
| Revenue from freight transportation | 228851.1 | 200803.4 | 215736.8 | 157968.2 | 250819 | 241591.6 | 241070.5 | 301048.6 | 354568.7 | 357465.9 | 348847.3 | 359571.4 | 401152.3 | 296154.2 | 267062.6 |
| Among them: local | 26206.4 | 20735 | 16457.4 | 22412.1 | 23008.6 | 18296.8 | 13059.9 | 17218.9 | 24797.4 | 25035.1 | 24481.2 | 24408.5 | 28146.9 | 21791 | 21436.6 |
| Import | 10626.6 | 13705.3 | 17700.7 | 23187.5 | 31636.2 | 30260.6 | 31829.1 | 37158.2 | 40042.2 | 39681.4 | 36732.4 | 44922.8 | 53428.2 | 50446.5 | 53806.9 |
| Export | 11963 | 8737.1 | 10892 | 12434.3 | 12402.8 | 14189.4 | 21106.2 | 28490.1 | 27829.4 | 23681 | 24433.3 | 26312.6 | 27657.9 | 25504.9 | 27055.7 |
| Transit | 180055.1 | 157626 | 170686.7 | 199934.4 | 183771.4 | 178844.8 | 175075.4 | 218181.5 | 261899.7 | 269068.4 | 263200.4 | 263927.5 | 291919.3 | 198411.8 | 164763.4 |

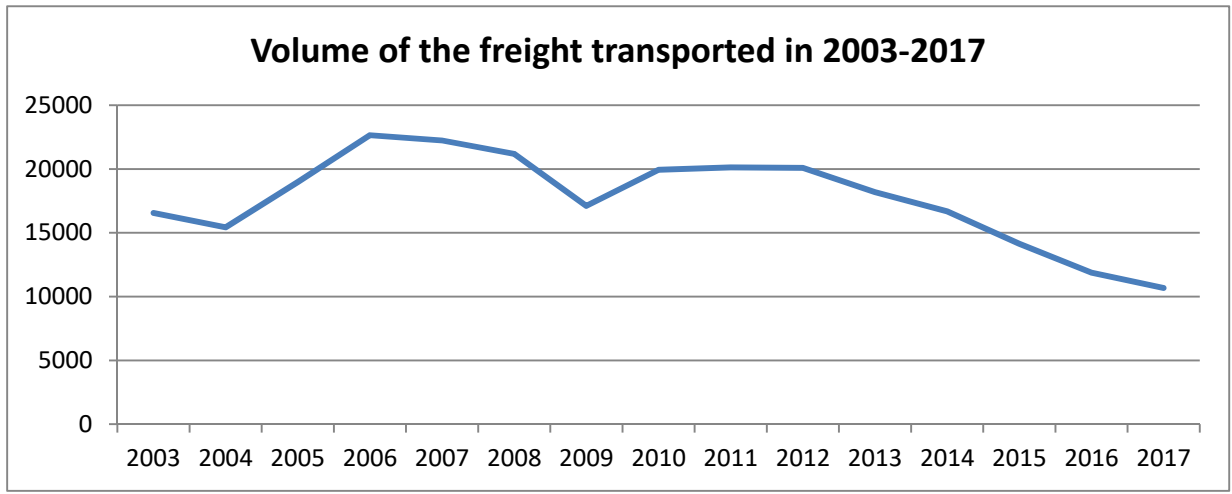

Figure 1. Volume of the freight transported in 2003-2017

Source: The figure is built from the Table2 Table. The horizontal axis shows the years, and vertical quantities - the volume of shipped cargo.

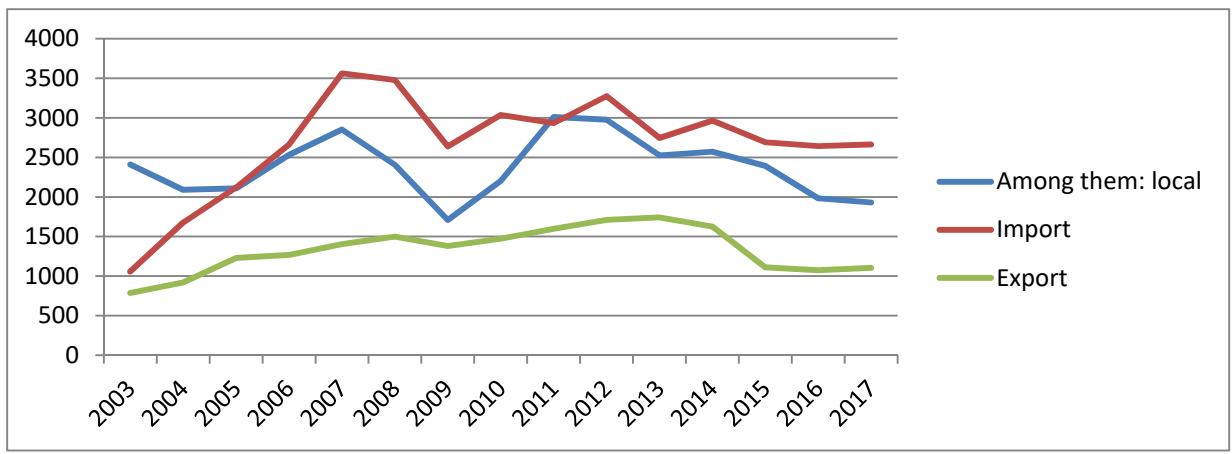

Figure 2. Railway transportation in 2003-2017 - including local, import, export

Source: The figure is built from the Table2 Table. The horizontal axis shows the years, and vertical quantities - the volume of shipped cargo.

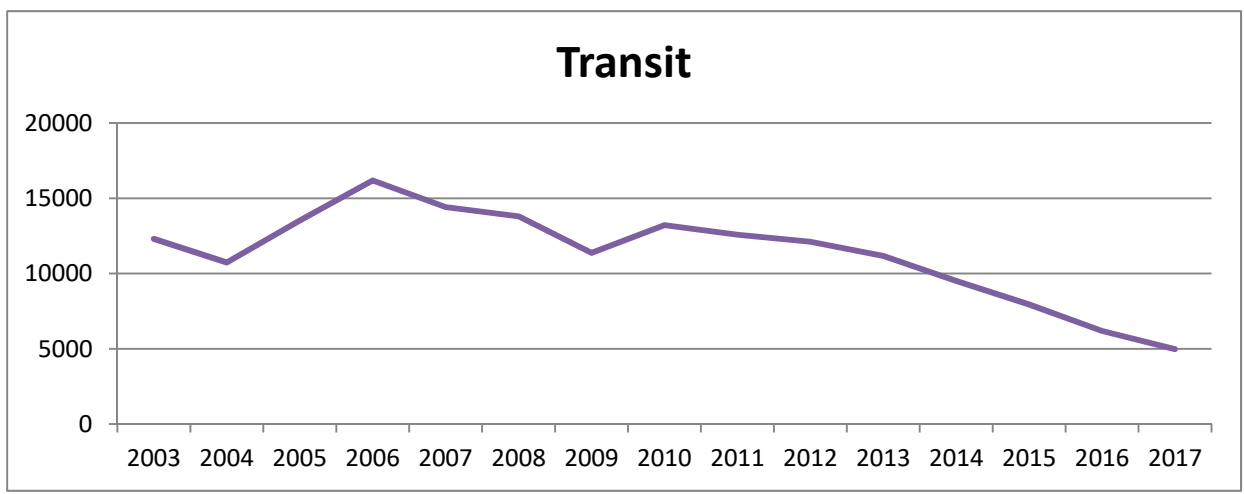

Figure 3. Transit railway transfers in 2003-2017

Source: The figure is built from the Table2 Table. The horizontal axis shows the years, and vertical quantities - the volume of shipped cargo.

## 4. Research Results

using the research methods and the findings of the internal and external factors we can make the assumptions regarding the optimistic scenario that, after the full modernization of the Georgian Railway`s Main Railway line which will give the company opportunity to increase the freight transportation capacity from existing 26 million tons per year up to 50 million tons per annual. In addition to above mentioned, as a result of the development of the correct strategies for the freight transportation (taking into consideration external factors) we can expect the sharp increase of transit transportation carried out by the Georgian Railway JSC.

The matrix inputs are listed below:

1. Period 1-16 years;

2. Growth Mathematical Model - compound annual growth rate (CAGR (Note 9))

3. The average weighted tariff used as of today is 10 dollars per 1 km/tones and remains unchanged through modeling.

4. Inflation rate is not used in the analysis.

5. Correction of Shipping Costs under the Effective Management of the Company

6. Added Economic Value Model (EVA) is used for the calculation of the Economic results= Income - Cost - Used Assets on X Discount Rates, calculated by Capital Asset Pricing Model (CAPM);

7. The current value of the effect balance - is calculated by net present value (NPV) model

8. During the analysis, the research also uses Georgia`s measurement of economic growth GDP- gross domestic product, based on its growth rate which is currently 4%.

Table 3. Matrix (Results)

| Matrix (Results) Growth in Freight Transportation (CAGR %) | The total volume of transportation after 16 years | Cost adjustment % | GDP in current prices, mln. Gel | The current value of the effect balance | The economic share of railway in GDP |
| --- | --- | --- | --- | --- | --- |
| 1% | 12.41 | 1% | $530,161 | ($59.66) | -0.011% |
| 2% | 14.12 | 1% | $530,161 | ($19.68) | -0.004% |
| 3% | 15.84 | 1% | $530,161 | $20.30 | 0.004% |
| 4% | 17.55 | 1% | $530,161 | $60.28 | 0.011% |
| 5% | 19.26 | 1% | $530,161 | $100.26 | 0.019% |
| 6% | 20.97 | 1% | $530,161 | $140.25 | 0.026% |
| 7% | 22.68 | 1% | $530,161 | $180.23 | 0.034% |
| 8% | 24.40 | 1% | $530,161 | $220.21 | 0.042% |
| 9% | 26.11 | 1% | $530,161 | $260.19 | 0.049% |
| 10% | 27.82 | 1% | $530,161 | $300.17 | 0.057% |
| 11% | 29.53 | 1% | $530,161 | $340.15 | 0.064% |
| 12% | 31.24 | 1% | $530,161 | $380.13 | 0.072% |
| 13% | 32.96 | 1% | $530,161 | $420.11 | 0.079% |
| 14% | 34.67 | 1% | $530,161 | $460.09 | 0.087% |
| 15% | 36.38 | 1% | $530,161 | $500.08 | 0.094% |

We have chosen a system approach and first used (EVA, CAPM, NPV) in the financial-economic models to determine the economic value of the Georgian Railway, thus determining the railway sensitivity of the railway in the country's economy. Which show in Table 3, the first column of the matrix shows the average annual geometric growth (CAGR) and after each simulation is increased by 1%. The second column gives the result of the CAGR resulting on the total volume of shipments, starting from year 2017 (approximately 10,7 million tons). The third column provides the simulation where the cost is reduced by 1%. The fourth column provides the information regarding the current value of GDP received from 2017 (taking into account the annual 4% increase since 2017). The fifth column presents the current value of the added value created on the railway based on the previous combinations and assumptions, and finally, in the last column we can see the share of the economic added value of the railway industry in the overall GDP in GDP, meaning the share that will be added to the current value after the changes made on CAGR.(Note 10)

## 5. Conclusions

Based on the statistical analysis of the research, the following was established

1. The assessment of the railway transport`s performance is carried out by checking the level of implementation of quantitative and qualitative indicators. For that purpose, the article has analyzed the technical and economic indicators of the Georgian Railway`s performance for the period of 2003-2017 and consequently the diagrams are put together showing that the total volume of freight transported in 2017 (by taking into account the average annual geometric data CAGR since 2003) has been decreased by 3,09%. In addition to that, the volume of the local shipments has been decreased by 1,57%, the volume of the import was increased by 6,82%, the volume of the export increased by 2,45% and finally the volume of the transited cargo has been decreased by 6,26%.

2. The total revenue received from freight transportation during the years 2003-2017 was increased by 1,11% according to the average annual geometric data (CAGR). The total income from local shipments decreased by 1,42%, revenue from import increased by 12.28%, revenue from export increased by 6% and revenue from transit was reduced by 0.63%.

3. The economic share of the railway industry in country`s overall GDP in 2006-2017 has been decreased from 0.476% to 0.145%.

4. The Georgian Railways opportunities regarding the increase in the volume of transit transportation has been investigated and concluded that the main railway line of the Georgian Railway can carry not less than 30-35 million tons of freight per year in the conditions of the increased demand on the services.

The main findings of the research can be considered according to the Average compound annual geometric rate (CAGR) increase in cargo volume in 16-year cycle allows the "Georgian Railway" JSC to add value to the country's overall GDP by the following model: (Note 11)

- 3% Increase – the additional value to GDP 0.004%;
- 4% Increase – the additional value to GDP 0.011%;
- 5% Increase – the additional value to GDP 0.019%;
- 6% Increase – the additional value to GDP 0.026%;
- 7% Increase – the additional value to GDP 0.034%;
- 8% Increase – the additional value to GDP 0.042%;
- 9% Increase – the additional value to GDP 0.049%;
- 10% Increase – the additional value to GDP 0.057%;
- 11% Increase – the additional value to GDP 0.064%;
- 12% Increase – the additional value to GDP 0.072%;
- 13% Increase – the additional value to GDP 0.079%;
- 14% Increase – the additional value to GDP 0.087%;
- 15% Increase – the additional value to GDP 0.094%;

Rates are a simulation of the regression model which gives the result of the additional economic value (EVA) that creates the railway GDP with the CAGR model of cargo growth. Therefore, the interest rate mentioned in the regressive model is adopted by the following mathematical formula: (Railway PV of EVA) / (PV of FV GDP).

The findings of the research indicate that the additional value to GDP is the direct and indirect form of development and growth of various sectors of the economy of Georgia, as some part of the cargo shipped in the railway remains in Georgia and is used in the process of production, which in itself adds value added to the economic growth of the country. Also, the use of this model by scientific research in foreign research centers provides better opportunities for additional growth of their economies.

**Notes**

Note 1. National statistics office of Georgia. Gross Domestic Product at current prices by 45 activities http://www.geostat.ge/index.php?action=page&p_id=119&lang=eng

Note 2. National statistics office of Georgia. Foreign Direct Investments by Economic Sectors.

http://www.geostat.ge/index.php?action=page&p_id=2231&lang=eng

Note 3. https://www.globalrailwayreview.com/article/71309/railroad-freight-gdp-growth/

Note 4. What is the contribution of rail to the UK economy? Prepared for the Rail Delivery Group July 2014 https://www.oxera.com/wp-content/uploads/2018/07/Contribution-of-rail-to-the-UK-economy-140714.pdf.pdf

Note 5. The Georgian Railway is an infrastructural company designed to transport cargo from Central Asia to Europe and vice versa. When these regions are observed to reduce economic activity, this will be reflected in the shipment of cargo from one point to another and vice versa, which in itself reflects the sensitivity of the railway turnover.

Note 6. Economic value added (EVA) is a measure of a company's financial performance based on the residual wealth calculated by deducting its cost of capital from its operating profit, adjusted for taxes on a cash basis. Note 7. EVA can also be referred to as economic profit, as it attempts to capture the true economic profit of a company. https://www.investopedia.com/terms/e/eva.asp

Note 8. Net present value (NPV) is the difference between the present value of cash inflows and the present value of cash outflows over a period of time. NPV is used in capital budgeting and investment planning to analyze the profitability of a projected investment or project. https://www.investopedia.com/terms/n/npv.asp

Note 9. The Capital Asset Pricing Model (CAPM) describes the relationship between systematic risk and expected return for assets, particularly stocks. CAPM is widely used throughout finance for pricing risky securities and generating expected returns for assets given the risk of those assets and cost of capital. www.investopedia.com/terms/c/capm.asp

Note 10. The compound annual growth rate (CAGR) is the rate of return that would be required for an investment to grow from its beginning balance to its ending balance, assuming the profits were reinvested at the end of each year of the investment's lifespan. https://www.investopedia.com/terms/c/cagr.asp

Note 11. The initial data of GDP (from where it is simulated in the regressive model - increase and then its current value) is represented on the link http://www.geostat.ge/index.php?action=page&p_id=119&lang=geo

The presented interest rates are a simulation of the regression model which gives the result of the additional economic value (EVA) that creates the railway GDP with the CAGR model of cargo growth. Therefore, the interest rate mentioned in the regressive model is adopted by the following mathematical formula: (Railway PV of EVA) / (PV of FV GDP).